\documentclass{article}[12pt]
\usepackage{amsmath}
\usepackage{amssymb}
\usepackage{amsthm}
\usepackage{amsfonts}
\usepackage[dvips]{graphicx}

\def\abstract#1{\vskip 7mm
        \begin{center}{\large Abstract}\par \smallskip
                \begin{minipage}[c]{12cm}
                        \small #1
                \end{minipage}
        \end{center}
}

\def\title#1{\begin{center}{\Large\bf #1}\end{center}}
\def\author#1{\vskip 5mm \begin{center}{#1}\end{center}}
\def\address#1{\begin{center}{\it #1}\end{center}}
\newcommand{\ssmatrix}[4]%
{\begin{pmatrix} #1 & #2 \\ #3 & #4 \end{pmatrix}}
\makeatletter
\@ifundefined{lesssim}{}{}
\@ifundefined{gtrsim}{}{}
\def\vereq#1#2{\lower3pt\vbox{\baselineskip1.5pt \lineskip1.5pt
\ialign{$\m@th#1\hfill##\hfil$\crcr#2\crcr\sim\crcr}}}
\makeatother
\renewcommand{\O}{\Omega}
\renewcommand{\o}{\omega}
\newcommand{\ti}{\tilde}
\renewcommand{\sc}{\mathscr{I}^+}
\newcommand{\scc}{\mathscr{I}^-}

\newcommand{\ri}{\rightarrow}
\newcommand{\bq}{\begin{equation}}
\newcommand{\ee}{\end{equation}}
\newcommand{\f}{\frac}

\newcommand{\bee}{\begin{equation}}
\newcommand{\eee}{\end{equation}}
\newcommand{\beq}{\begin{eqnarray}}
\newcommand{\eeq}{\end{eqnarray}}
\newcommand{\na}{\tilde{\nabla}}
\newcommand{\pa}{\partial}
\newcommand{\sq}{\sqrt}

\newcommand{\bc}{\begin{center}}
\newcommand{\ec}{\end{center}}
\newcommand{\bl}{\biggl}
\newcommand{\br}{\biggr}
\newcommand{\beqq}{\begin{eqnarray*}}

\newcommand{\no}{\nonumber}

\newtheorem{definition}{Definition}

\usepackage{ascmac}
\usepackage{rsfs}

\begin{document}
\setlength{\baselineskip}{16pt}

\title{Limit structure of Future Null Infinity tangent\\
-topology of the event horizon and gravitational wave tail-}
\author{Shinya Tomizawa
\footnote{tomizawa@sci.osaka-cu.ac.jp}
 and Masaru Siino
\footnote{msiino@th.phys.titech.ac.jp}}
\address{Department of Physics, Osaka City University, 3-3-138, Sugimoto, 
Sumiyoshiku, Osaka City, Osaka, 113-0033, Japan}
\address{Department of Physics, Tokyo Institute of Technology, Oh-Okayama, Tokyo 152-8550, Japan}
\date{\today}

\abstract{
We investigated the relation between the behavior of gravitational wave at late time and the limit structure of future null infinity tangent which will determine the topology of the event horizon far in the future. 
In the present article, we mainly consider a spacetime with two black holes. 
Although in most of cases, the black holes coalesce and its event horizon is topologically a single sphere far in the future, there are several possibilities that the black holes never coalesce and such exact solutions as examples. 
In our formulation, the tangent vector of future null infinity is, under conformal embedding, related to the number of
black holes far in the future through the Poincar\'e-Hopf's theorem. 
Under the conformal embedding, the topology of event horizon far in the future will be affected by the geometrical structure of the future null infinity. 
In this article, we related the behavior of Weyl curvature to this limit behavior of the generator vector of the future 
null infinity. 
We show if Weyl curvature decays sufficiently slowly at late time in the neighborhood of future null infinity, two black holes never coalesce.
}

%\keywords{future null infinity; asymptotically flat; topology}

%\maketitle

\section{\label{sec:1}Introduction}

We are often interested in the final state of a black hole spacetime after gravitational collapse. 
When matter collapses, it passes inside an event horizon and energy will be emitted to future null infinity in the form of gravitational waves. 
Since the amount of this energy is limited, we may expect that the spacetime will approach to a stationary state. If a black hole spacetime becomes stationary far in the future to an observer outside the black hole, the final state of the black holes is considerably restricted. 
As used in reference \cite{Siino}, the term {\it far in the future} means {\it sufficiently at late time to an observer which remains outside a black hole}.

So far the final state of the black hole spacetime has been studied by many authors and these studies are known as uniqueness theorems of black holes. Israel showed that the only static and topologically spherical black hole is the Schwarzschild solution \cite{Israel} or the Reissner-Nordstr\"{o}m solution \cite{Israel2}. 
In stationary axisymmetric situations, the uniqueness theorems of a vacuum black hole and a charged black hole were shown by Carter \cite{Carter}, Robinson \cite{Robinson},\cite{Robinson2} and Mazur \cite{M}. 

All they are based on the topological notion, 
which is also well studied for stationary black holes\cite{Hawking}\cite{Chrusciel and Wald}\cite{OW}.
Hawking proved that the spatial topology of the smooth stationary horizon must be spherical\cite{Hawking}.
Moreover, the work of Chru\'sciel and Wald\cite{Chrusciel and Wald} showed that in a stationary spacetime,
the spatial topologies of the connected components of black holes are only spheres under null energy condition.

Hence, now it
is known that, under some reasonable conditions such as asymptotic
flatness and the weak energy condition, 
each component of the black hole region is topologically trivial,
i.e., simply connected\cite{Chrusciel and Wald}.
On the other hand, there were numerical simulations which suggest
non-trivial topologies of the horizons\cite{ONW}. 
There have been some confusion, but it is now well understood that 
even though the black hole region in the spacetime is simply connected, 
there are many possible topologies of spatial sections. 
In particular, 
one of the authors\cite{Siino} showed how topology of the 
spatial sections of a black hole is related to 
the endpoint set, or similarly, the {\em crease set}, of the event 
horizon. 

Thus, we understand that the topology of the event horizon can be realized by the formation stage with the crease set
and the final stationary stage, independently. 
If the black hole spacetime settles down to a stationary state, 
we expect the spatial topology of the event horizon will consist of some connected components with spherical topology. 
To realize these aspects more, a new approach is required to investigate the topology of an event horizon far in the future. 
Especially the number of connected components of the black hole is most important.

There is not a simple method to know topological structure of an event horizon far in the future, that is, to determine how many black holes the final state of the black hole spacetime will consist of. 
Then we consider some examples, 
they are Majumdar-Papapetrou solution\cite{Hartle and Hawking}, C-metric\cite{Pravda} (which
recently becomes important in studies of higher dimensional black holes since the black ring solution \cite{br} with 
vanishing angular momentum corresponds to a higher dimensional C-metric) and so on. 
The spatial topology of the event horizon of such a solution is obviously not a single sphere.

Since in an asymptotically flat spacetime\cite{Wald}, (to be more exact, in a strongly asymptotically predictable 
spacetime) an event horizon is defined as the boundary of the causal past of future null infinity $\sc$, 
we expect the topology of an event horizon will be strongly related to the topological structure of future null infinity. 
Because the topology of future null infinity in an asymptotically flat spacetime is $S^2\times {\bf R}$, 
we might conclude far
 in the future the spatial topology of an event horizon will always be a single sphere. 
Nevertheless it is well known that there are several above mentioned stationary solutions including black holes
separated from each other.

Therefore we have to study the structure of the neighborhood of $\sc$ in detail, especially in the limit of the
future direction.
The geometric study of the asymptotic region was successful in the formulation of large distance expansion
 (the expansion of a conformal factor $\Omega$ of conformal embedding) like the investigation of the peeling property\cite{pp}.
Then we evolve the large distance expansion of geometry into the future direction along the tangent
of the future null infinity $n^a=\tilde{g}^{ab}\tilde{\nabla}_b\Omega$.

To analyse that, a key aspect is that there remains a long tail of Weyl curvature $\Psi_{4}$ in the C-metric spacetime,
which is discovered by Tomimatsu\cite{Tomimatsu}.
In the present article, we develop pseudo-Newman Penrose basis vectors and clarify
the relation between the Weyl curvature component of outgoing gravitational wave and 
the congruence of the integral curve of
the $\sc$ tangent vector $n^a$.
We will find these aspects rule the topology of an event horizon in the limit of those integral curves.

To determine the gauge condition of the conformal embedding for studying this limit structure of the tangent
field $n^a$, we should be careful since such a limit will be realized
only when conformal embedding makes unphysical manifold compact in this direction.
We show from the discussion in such a unphysical manifold that two black holes are separated forever for an asymptotic 
observer outside black holes, if its Weyl curvature in the neighborhood of future null infinity decays slowly at late time.     

 Since we discuss black hole spacetimes, only asymptotically flat spacetime is considered. 
So, in section \ref{sec:2}, we will give the notion of an asymptotically flat spacetime and the gauge condition of conformal
embedding.
The congruence of the integral curves of tangent field $n^a=\tilde{g}^{ab}\tilde{\nabla}_b\Omega$ is investigated in the third section.
Fourth section provides the index theorem and using it, study the relation between the topology of an event horizon 
and the structure of that congruence.
In section \ref{sec:5}, we will apply this discussion to C-metric as an example, to illustrate its detail concretely.  
The final section is devoted to summary and discussions.
Throughout this article, we use the abstract index notation as the component notation of tensors and it is denoted by Latin indices $a,b,\cdots$.

\section{\label{sec:2} Asymptotic flatness and gauge choicing of conformal compactification}

In this article, we use the term {\it far in the future} to mean {\it at sufficiently late time to an observer outside black holes} in stead of referring the time after a spacetime settles down to a stationary phase. Then, our main question is ``How can we know the topology of an event horizon far in the future ?"

To discuss the topology of an event horizon, we pay attention to the relation with
future null infinitiy since the event horizon is defined as the boundary of the past set of future null infinity $H^+=\partial(J^-(\sc))\cap M$.
Because of its non-compactness ($\sc\sim S^2\times {\bf R}$)\cite{Wald} there is an ambiguity of the limit structure.

Here our tactics is to compactify the future null infinity and the integral curves of $n^a$ tangent to $\sc$ in their 
direction to discuss the limit of the integral curves.
Therefore inextendible endpoints of the integral curve of $n^a$ will be attached after gauge choice of the conformal embedding.
It will be discussed that a geometric condition to allow that a set of endpoints with non-trivial topology is attached, by the Poincar\'{e}-Hopf theorem.
From this we discuss the topology of an event horizon far in the future.

First of all, we mention asymptotic flatness.
Asymptotic flatness is essential in our discussion.
 We have to state the definition of asymptotically flat spacetime before explaining the analytic method to examine the limit structure of $\sc$.  The definition mentioned here is based on the reference \cite{Wald}.

\begin{definition}[Asymptotic Flatness]
A spacetime $(M,g_{ab})$ is said to be asymptotically flat at null infinity if there exists an unphysical spacetime $(\tilde{M},\tilde{g}_{ab})$ with $\tilde{g}_{ab}=\O^2g_{ab}$ smooth everywhere except at the point $i^0$ where it is $C^{>0}$ and conformal isometry $\psi:M\rightarrow\psi (M)\in\tilde{M}$ with conformal factor $\Omega$ satisfying the below conditions. 
\begin{enumerate}
\item $\bar{J}^+(i^0)\cup\bar{J}^-(i^0)=\tilde{M}-M$. Thus, $i^0$ is spacelike related to all points in $M$ and the boundary of $M$ consists of $i^0$, $\sc\equiv\dot{J}^+(i^0)-i^0$ and $\scc\equiv\dot{J}^-(i^0)-i^0$.
\item There exists an open neighborhood $V$ of $\dot{M}=i^0\cup\sc\cup\scc$ such that the spacetime $(V,\tilde{g}_{ab})$ is strongly causal.
\item $\Omega$ can be extended to a function on all $\tilde{M}$ which is $C^{\infty}$ everywhere and  $C^{>0}$ at $i^0$.
\item On $\sc$ and $\scc$ we have $\O=0$ and $\na_a\O\ne 0$.
\item The map of null direction at $i^0$ into the space of integral curves of $n^a\equiv\tilde{g}^{ab}\tilde{\nabla}_b\Omega$ on $\sc$ and $\scc$ is diffeomorphism. 
\item For smooth function $\o$ on $\ti{M}-i^0$ with $\o>0$ on $\ti{M}\cup\sc\cup\scc$ which satisfies $\tilde{\nabla}_a(\omega^4n^a)=0$ on $\mathscr{I}^+$, the vector field $\omega^{-1}n^a$ is complete on $\sc$ and $\scc$.
\item In the neighborhood of $\sc$ and $\scc$ physical Ricci tensor behaviors as $R_{ab}=O(\Omega^2)$.
\end{enumerate}
\label{def:af}
\end{definition}

Let us note that there is gauge freedom in the choice of an unphysical spacetime $(\tilde{M},\tilde{g}_{ab})$ with an asymptotically flat physical spacetime $(M,g_{ab})$. This gauge freedom is important in our discussion.
 
\textit{If the spacetime $(\tilde{M},\tilde{g}_{ab})$ is an unphysical spacetime satisfying the above definition with conformal factor $\O$, so is $(\tilde{M},\omega^2\tilde{g}_{ab})$ with conformal factor $\o\O$ for the any function $\omega$ which is smooth everywhere except at $i^0$ and positive everywhere.}

The most common and useful gauge condition will be an {\it affine gauge}. 
In this gauge, the tangent $n^a=\tilde{g}^{ab}\tilde{\nabla}_b\Omega$ of $\sc$ generators is affinely parametrized. See the equation (11.1.21) in reference \cite{Wald}. We should note that in this gauge, the null generators $n^a$ (so, the future null infinity generated out of them) are not compactified in the future direction, since the integral curves of $n^a$ have infinite ranges of affine parameters in the future direction.  
Nevertheless in the present article, we consider another gauge condition such that the generator of $\sc$ can be 
compactified.
Our purpose is to investigate the congruence of the $n^a$-integral curves around the future null infinity in the limit of their parameter.
To study this, it will be valid to compactify the future null infinity and the integral curves in their direction.
The compactification will be realized by attaching any endpoints of the $n^a$-integral curves, as $\sc$ has been attached to
 $M$ to study its asymptotic structure.
Therefore $\omega$ vanishes at the set $\cal E$ of attached endpoints.
Taking a sufficiently small neighborhood of $\cal E$, $\omega$ is considered to be monotonically decreasing in the direction of $n^a$-integral curves.
We call this gauge a {\it compact gauge}.

Here it should be commented that according to the gauge condition, the integral curves go to $\sc$, $i^+$ or singularity.

\section{\label{sec:4}Limit structure of future null infinity}
We investigate the geometrical structure around future null infinity in a limit
of the integral curves of $n^a=\tilde{g}^{ab}\tilde{\nabla}_b\Omega$. 
We define the limit as the limit approaching to the attached endpoints $\cal E$.
We call this limit $n^a$-limit.

\subsection{Physical Ricci Tensor and Unphysical Ricci Tensor}
The relation between the physical Ricci tensor and the unphysical Ricci tensor gives us a great deal of information. 
The physical Ricci tensor $R_{ab}$ is related to the conformally translated unphysical Ricci tensor $\tilde{R}_{ab}$ by

\beq
R_{ab}&=&\ti{R}_{ab}+2\O^{-1}\na_a\na_b\O \no\\
      &{}&+\ti{g}_{ab}\ti{g}^{cd}(\O^{-1}\na_c\na_d\O -3\O^{-2}\na_c\O\na_d\O).\label{eq:R}
\eeq

By multiplying $\O^2$ and taking the limit $\O\ri 0$ we find that a vector $n^a\equiv\ti{g}^{ab}\na_b\O$ must be extended smoothly to $\sc$ and be null at $\sc$ because the first term vanishes from conditions 4 and 7 of the definition \ref{def:af}, the second vanishes from conditions 3 and 4, and the fourth vanishes from the condition 3,4 and $\ti{g}_{ab}$ being smooth at $\sc$.

$n^a$ is null at $\sc$ but is not null off of $\sc$ and we put the shift from null vector as the following

\begin{equation}
\ti{g}_{ab}n^an^b=k_{(1)}\O+k_{(2)}\O^2+\cdots ,\label{eq:nn}
\end{equation}
where $k_{(i)}\ (i=1,2,\cdots)$ are the functions independent of $\O$. 
The integral curves of $n^a$ on $\sc$ is null geodesic generators of $\sc$, but the integral curves of $n^a$ is not geodesics off of $\sc$. Let us put the shift from the geodesic as
\begin{equation}
n^a\na_an^b-p_{(0)}n^b=O(\O).
\end{equation} 
In particular, we put the contraction of the above equation with $l^a$ as the following:
\begin{equation}
l^bn^a\na_an_b=-p_{(0)}+p_{(1)}\O+\cdots ,\label{eq:lnn}
\end{equation}
where we determine the normalization as $l^an_a=-1$ (see the below equation (\ref{eq:norln})).

From (\ref{eq:nn}) and (\ref{eq:lnn}), we obtain the relation between $k_{(i)}$ and $p_{(i)}$
\begin{equation}
p_{(0)}=\f{1}{2}k_{(1)},\ p_{(1)}=-k_{(2)},\ p_{(2)}=-\f{3}{2}k_{(3)},\cdots ,\label{eq:pk}
\end{equation}
where we used the relation of torsion free ;
\begin{equation}
\na_an_b=\na_bn_a .
\end{equation}

\subsection{Pseudo-Newman-Penrose Formalism}
Newman-Penrose formalism~\cite{Newman Penrose} is the choice of the null basis which consists of a pair of real null vectors, $l^a, n^a$ and a pair of complex conjugate null vectors $m^a, \bar{m}^a$. 
Now we consider $n^a\equiv \ti{g}^{ab}\na_b\O$ as one of real null vectors in Newman-Penrose formalism. 
As we have mentioned before, $n^a$ is null on $\sc$ but is not null away off $\sc$. 
Therefore, strictly speaking, this is not Newman-Penrose basis vectors. 
Nevertheless most of Newmann Penrose framework are not affected by this difference.

So the following orthogonality conditions are required
\beq
l^am_a=l^a\bar{m}_a=n^am_a=n^a\bar{m}_a=0\label{eq:orthogonal}
\eeq 
in addition to the conditions that vectors are null,

\begin{equation}
l^al_a=m^am_a=\bar{m}^a\bar{m}_a=0,
\end{equation}
and the fact that $n^a\equiv\ti{g}^{ab}\na_b\O$ is null only on $\sc$,

\begin{equation}
n^an_a=k_{(1)}\O+k_{(2)}\O^2+\cdots.\label{eq:nn2}
\end{equation}
We impose on the basis vectors the further normalization conditions,
\begin{equation}
l^an_a=-1,\ m^a\bar{m}_a=1,\label{eq:norln}
\end{equation}
which is consistent with the definition of the vector $n^a$ and $l^a$.
Thus, the metric can be represented by 
\begin{equation}
\ti{g}_{ab}=-(k_{(1)}\O+k_{(2)}\O^2+\cdots)l_al_b-2l_{(a}n_{b)}+2m_{(a}\bar{m}_{b)}\end{equation}
in this pseudo-Newman-Penrose basis vectors. Note that the first extra term of the above equation is caused by the equation (\ref{eq:nn2}).

In order to investigate the congruence of the $n^a$-integral curves around $\sc$, we have only to consider the values of spin connections and curvatures only in the neighborhood of $\sc$. 
Now we expand the each basis component of the equation (\ref{eq:R}) in the powers of conformal factor $\O$ as $x=x_{(0)}+x_{(1)}\Omega+x_{(2)}\Omega^2+...$.  
The $(n,l)$ component of the equation (\ref{eq:R}) gives
\beq
O(\O^2) &=& \bl(\f{5}{2}k_{(1)}-3p_{(0)}+2\mu_{(0)}\br)\O^{-1}\no\\
  &{}& +\bl(\ti{R}^{(0)}_{nl}+3p_{(1)}+2k_{(2)}+2\mu_{(1)}+k_{(1)}q\br)+O(\O)\label{eq:rev}\\
\eeq
where we used the condition 7 of the above definition \ref{def:af} and the equations (\ref{eq:nn}),(\ref{eq:lnn}) and $\mu\equiv-m^a\bar{m}^b\na_an_b$ is the spin connection, which means the expansion of the integral curves of $n^a$. $q$ is the function satisfying $l^a\na_al^b=ql^b$.

So we obtain from the coefficients of the first and the second term 
\begin{equation}
\f{5}{2}k_{(1)}-3p_{(0)}+2\mu_{(0)}=0\label{eq:mpk}
\end{equation}
and
\begin{equation}
\ti{R}^{(0)}_{nl}+3p_{(1)}+2k_{(2)}+2\mu_{(1)}+k_{(1)}q=0.
\end{equation}

Similarly, from other components we can obtain the relations between spin connections and Ricci tensors.

From the $(l,l)$ component, we obtain 
\begin{equation}
q=0,
\end{equation}
\begin{equation}
\ti{R}_{ll}^{(0)}=0.
\end{equation}
From the $(m,m)$ component, we obtain
\begin{equation}
\bar{\lambda}_{(0)}=\lambda_{(0)}=0,\label{eq:lam}
\end{equation}
\begin{equation}
\lambda_{(1)}=\ti{R}_{mm}^{(0)},
\end{equation}
where $\lambda\equiv -\bar{m}^a\bar{m}^b\na_an_b$ is the spin connection which means the shear of the integral curves of vector field $n^a$.
Eq.\ref{eq:lam} means $\sc$ ($\Omega\equiv 0$) is diffeomophic to $S^2\times{\bf R}$.
The $(n,m)$ component gives
\begin{equation}
\nu_{(0)}=\bar{\nu}_{(0)}=0,
\end{equation}
\begin{equation}
2\nu_{(1)}=\ti{R}_{nm}^{(0)},
\end{equation}
where $\nu$ is defined by $\nu\equiv n^a\bar{m}^b\na_an_b$. 

\subsection{Change of Conformal Factor}
Now it should be recalled that there is gauge freedom in the conformal transformation considered above. 
Under the transformation $\O\ri\O'=\o\O, \ \ti{g}_{ab}\ri\ti{g}'_{ab}=\o^2\ti{g}_{ab}$, we have

\bq
n^a\ri n^{\prime a}=\o^{-1}n^a+\O\o^{-2}\ti{g}^{ab}\na_b\o.
\end{equation}
In particular, on $\sc$ the above transformation becomes
\begin{equation}
n^a\ri n^{\prime a}=\o^{-1}n^a.
\end{equation}

Under this transformation, the expansion of the null geodesic generators is transformed as
\begin{equation}
\mu_{(0)}\ri\mu'_{(0)}=\f{\mu_{(0)}}{\o}-\f{1}{\o^2}n^a\na_a\o
\label{eq:mm}
\end{equation}
on $\sc$.
Therefore, note that we can choose $\mu_{(0)}$ as any function independent of $\O$.  
Given $\mu'_{(0)}$, since  the equation (\ref{eq:mm}) is merely an ordinary differential equation, there always exists $\o$ satisfying the equation (\ref{eq:mm}). 
From the equations (\ref{eq:pk})(\ref{eq:mpk}), we find the relation
\begin{equation}
\mu_{(0)}=-p_{(0)}=-\f{1}{2}k_{(1)}.\label{eq:mupk}
\end{equation}
So from the equation (\ref{eq:lnn}), we see that in the gauge satisfying $\mu_{(0)}=0$, the null geodesics on $\sc$ are affinely parameterized. 
In this article, we call this gauge \textit{affine gauge}. 
When we discuss the limit structure around $\sc$, we should not choose the affne gauge since in the affine gauge, we cannot compactify $\sc$ or $n^a$-integral curve and is not convenient to study the congruence of $n^a$-integral curves in the $n^a$-limit on neighborhood of $\sc$. 
Hence, we choose the compact gauge such that $\sc$ can be compactified by attaching a set $\cal E$ of future endpoints of $n^a$, that is, $\o\ri 0$ in $n^a$-limit.

From (\ref{eq:lam}), the leading order for the shear $\lambda$, vanishes, which does not depend on the choice of the gauge; 
\begin{equation}
\lambda'_{(0)}=\lambda_{(0)}=0
\label{eqn:gtlm0}
\end{equation} 
holds.
Then also in this gauge, spatial section of $\sc$ ($\Omega\equiv 0$) is always $S^2$. 
$\sc$ is topologically $S^2\times(R\cup\{+\infty\})$ after future endpoints of its generators are attached.
Under the transformation $\O\ri\O'=\o\O$, the first order is transformed as
\begin{equation}
\lambda_{(1)}\ri\lambda'_{(1)}=\f{\lambda_{(1)}}{\o^2}-\f{1}{\o^3}\bar{m}^a\bar{m}^b\na_a\na_b\o-\f{2}{\o^4}(\bar{m}^a\na_a\o)^2.\label{eq:lamlam}
\end{equation}
From now on, $'$ represents the quantity after transforming from an affine gauge into a compact gauge.

\subsection{Congruence of $n^a$-curves and Weyl Curvature}
%As mentioned before, the topology of the event horizon is expected to be related to the upper end of $\sc$. We discuss this in the context of asymptotic flatness. In an asymptotic flat spacetime, it is guaranteed that there exists a conformal embedding defined in the latter half of section \ref{sec:2}. The asymptotic flatness, however, accepts the further conformal transformation that is indicated by $\omega$ as gauge freedom. Since the gauge transformation can become singular ($\omega$ can become zero or infinity) at the upper end of $\sc$, we should be careful to choose $\omega$. 

The gauge freedom can makes the congruence of $n^a$-integral curves shrink into a point or take away to infinity in the direction of $n^a$.
Indeed, when we take an affine gauge the null generators are complete in affine parameterization by definition and go away to infinity. 
Nevertheless, since we want to study the geometrical structure around $\sc$ in $n^a$-limit, it is essential to choose a compact gauge $\Omega'=\omega\Omega$, in which the $n^a$-integral curves are incomplete in affine parameterization (possible to compactify) though is complete in a original parameterization $n^a=(\partial/\partial u')^a$. 
Under this gauge condition, we discuss the limit structure of the congruence of the $n^a$-integral curves in $n^a$-limit.

When we choose the compact gauge we should be careful about the angular dependence of $\omega$, since we include the case of vanishing $\omega$.
If we allow $\omega$ to have angular-depending irregularity in $n^a$-limit, the irregular angular dependence may change the topological structure of the congruence of $n^a$-curve. 
Because a conformal transformation should be diffeomorphism, we must choose such $\omega$ that is smooth with regard to the angular and positive. 

For example, let us consider a sphere. 
The metric on a sphere with unit radius is $ds^2=d\theta^2+\sin^2\theta d\phi^2$. 
When we perform  the conformal transformation such that a conformal factor vanishes at $\theta=\pi/2$ (for example, $\omega=\cos\theta$, its metric is $ds^2=\cos^2\theta (d\theta^2+\sin^2\theta d\phi^2)$ and is not diffeomorphic to the original sphere. We must choose a regular gauge to avoid such a case. Then we never allow that dependence for the gauge transformation $\Omega'=\omega\Omega$ from an original affine gauge. 

Newman-Penrose equation\cite{Newman Penrose} relates the shear of the congruence around $\sc$ to Weyl curvature as the following;
\beq
-n^a\na_a\lambda+\bar{m}^a\na_a\nu&=&-(\mu+\bar{\mu})\lambda-(3\gamma-\bar{\gamma})\lambda\no\\
                                  &+&(3\alpha+\bar{\beta}+\pi-\bar{\tau})\nu+\Psi_{(4)}.\label{eq:NP}
\eeq

We expand each term of the above equation in powers of $\O$ and leave the only leading term. 
From (\ref{eq:lam}) the order of $\lambda$ is
\begin{equation}
\lambda=O(\O).\label{eq:lambda}
\end{equation}
The spin connection $\nu$ can be written as
\begin{equation}
\nu=-\f{1}{2}m^a\na_ak_{(1)}\O+O(\O^2)
\end{equation} 
by the equations (\ref{eq:nn}) and (\ref{eq:orthogonal}). We can choose $\o$ satisfying
\begin{equation}
m^a\na_ak_{(1)}=0\label{eq:omega}
\end{equation}
from the relation (\ref{eq:mupk}). 
For we can write $m^a$ as $m^a=m^{\theta}(\pa/\pa\theta)^a+im^{\phi}(\pa/\pa\phi)^a$ using two spacelike vectors, $(\pa/\pa\theta)^a, (\pa/\pa\phi)^a$ orthogonal to $n^a, l^a$. Then if we transform $\omega$ from affine gauge into another gauge, from the equation (\ref{eq:mm}), the equation (\ref{eq:omega}) becomes
\beq
\f{\pa}{\pa\theta}\biggl(\f{n^a\na_a\omega}{\omega^2}\biggr)=\f{\pa}{\pa\phi}\biggl(\f{n^a\na_a\omega}{\omega^2}\biggr)=0.
\eeq
We may choose $\omega$ so that $n^a\na _a\o /\o^2$ does not depend on $\theta,\phi$ and $\o$ is positive everywhere.  
Therefore, we see that for $\o$ satisfying (\ref{eq:omega}), the order of $\nu$ is
\begin{equation}
\nu=O(\O^2).\label{eq:nu}
\end{equation}

Finally, let us compute the order of the spin connection $\gamma$ defined by $\gamma\equiv 1/2(n^an^b\na_al_b-n^a\bar{m}^b\na_al_b) $. The real part of $\gamma$ is
\beq
\gamma+\bar{\gamma}&=&n^an^b\na_al_b\no\\
                   &=&\mu_{(0)}+O(\O)\label{eq:gpg}
\eeq
from the equations (\ref{eq:lnn}) and (\ref{eq:mpk}).  Since the imaginary part of $\gamma$ becomes 
\begin{equation}
\gamma-\bar{\gamma}=-n^a\bar{m}^b\na_am_b,
\end{equation}
it depends on the direction of $m^a$ which we have not yet determined. Now let us determine the direction of $m^a$ as follows.
We determine the direction of $m^a$ so that on $\sc$, it will be parallelly transported along the null geodesic generators on $\sc$. That is,
\begin{equation}
n^a\na_am^b=0\mbox{ on $\sc$.}
\end{equation}   
Moreover in the direction away off $\sc$, it is parallelly transported along $-l^a$, that is,
\begin{equation}
l^a\na_am^b=0\ \mbox{ everywhere.}
\end{equation} 
Thus, $\gamma-\bar{\gamma}$ becomes
\begin{equation}
\gamma-\bar{\gamma}=O(\O).\label{eq:gmg}
\end{equation}

So we see that from the equations (\ref{eq:gpg}) and (\ref{eq:gmg}), the order of $\gamma$ is
\begin{equation}
\gamma=\f{\mu_{(0)}}{2}+O(\O).\label{eq:gamma}
\end{equation}
From (\ref{eq:lambda}), (\ref{eq:nu}) and (\ref{eq:gamma}), we can write the equation (\ref{eq:NP}) as 
\begin{equation}
n^a\na_a\lambda_{(1)}=3\mu_{(0)}\lambda_{(1)}-\Psi_{4(1)}
\end{equation} 
in the leading order. 

Integrating this, we obtain the relation
\begin{equation}
\lambda_{(1)}=-\exp \left(3\int^{u'}\mu_{(0)}(u'')du''\right)\cdot \int^{u'}du''\left[\Psi_{4(1)}(u'')\exp\bl(-3\int^{u''}\mu_{(0)}(u''')du'''\br)\right],\label{eq:shear}
\end{equation}
where $u'$ is the parameter of $n^{\prime a}$ and is defined by $n^{\prime a}\equiv(\pa/\pa u')^a$.

\section{Topology of Event Horizon}

Now we discuss the relation between the topology of event horizon and the $n^a$-limit structure of $n^a$-integral curve 
congruence.
Pseudo-Newman-Penrose basis provides the following timelike submanifold near $\sc$ but not on $\sc$.
In this region, since $\omega$ is monotonically decreasing 
function in the direction of $n^a$ supposing sufficiently large parameters, $n^a$ is timelike from (\ref{eq:nn2})(\ref{eq:mm})(\ref{eq:mupk}) and orthogonal to a spatial hypersurface ${\cal C}_e$ which is
defined by small constant $e$ as $\Omega=e\ll 1$.
From the smoothness of the pseudo-Newman-Penrose basis, a sufficiently large sphere ${\cal S}_e$ is defined on 
${\cal C}_e$ and spanned by $m^a$ and $\bar{m}^a$.

$n^{\prime a}=(\partial/\partial u')^a$ generates a one-parameter family of diffeomorphisms $\psi_{u'}$ on $M$, and we have 
a family of two dimensional surface ${\cal S}_e(u')=\psi_{u'}({\cal S}_e)$ and
a submanifold
\begin{equation}
{\cal W}_e=\{p\in \psi_{u'}({\cal S}_e)|u'\in (-\infty,\infty)\},
\end{equation}
where $\psi_0$ identically maps ${\cal S}_e$ onto itself.

\begin{figure}[htbp]
\begin{center}
\includegraphics[width=.60\linewidth]{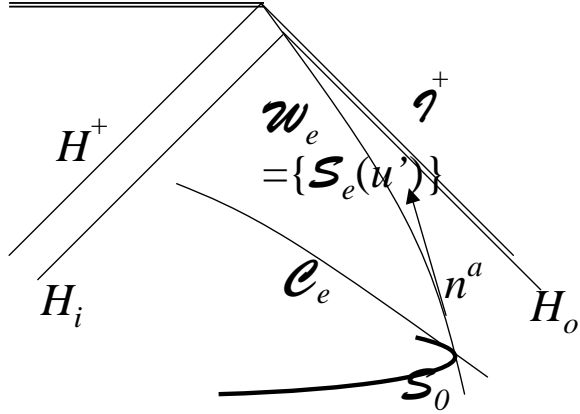}
\end{center}
\caption{${\cal C}_e$ is the spacelike surface where $\Omega$ is constant. On that a sufficiently large sphere ${\cal S}_e$
is given. $n^a$ generates a timelike submanifold ${\cal W}_e$ from it.
It determine a two boundaries of its past set $I^-({\cal W}_e)$. Inner one is $H_i$ and outer one is $H_o$.}
\label{fig:3}
\end{figure}

Since $\overline{M}$ is not a manifold at $i^+$, we want to discuss the structure of ${\cal S}_e(u')$ without dealing 
with $i^+$, that is within openset $I^-(\sc)$.
Nevertheless, the inextendible timelike curves on ${\cal W}_e$ does not have its endpoint on the future null infinity 
or $I^-(\sc)$, rather
on the outside of $I^-(\sc)$.
For sufficiently small $e$ and large ${\cal S}_e$, there is ${\cal W}_e$ outside of black hole, since the black hole
region $M\setminus I^-(\sc)$ is compact and $n^an_a=n^a\nabla_a\Omega =k_{(1)}\Omega+...<0$ means decreasing of $\Omega$ in $n^a$ direction.

Since ${\cal S}_e(0)=S_e$ is topologically a sphere, $I^-(\sc)$ is decomposed by ${\cal W}_e$ into an inside subset $A$ and an
outside subset $B$ (${\cal W}_e\subset A$).
If $( \dot{B} \cap M)\cap (\dot{J}^-(\sc)\cap M)\neq \emptyset$, ${\cal W}_e$ can cross the event horizon $H^+$.
Then 
\begin{eqnarray}
(\dot{B}\cap M)\cap (\dot{J}^-(\sc) \cap M)&=& \emptyset \\
 \dot{I}^-({\cal W}_e)&=&J^-(\sc)\cap M
\end{eqnarray}

The boundary of the causal past of ${\cal W}_e$ is composed of ingoing part $H_i=\overline{A}\setminus {\cal W}_e$ 
generated by ingoing
past null geodesics and outgoing part $H_o=\overline{B}\setminus {\cal W}_e$ generated by outgoing past null geodesics.
From the continuity of $M$, the ingoing part approaches to the event horizon. 
Considering $\lim_{u'\rightarrow\infty}{\cal S}_e(u')$, we may observe the event horizon in $n^a$-limit $H^+\cap U_E$ is
possibly not homeomorphic to the cross section of $\sc$, where $U_E$ is neighborhood of $E$ in $\overline{M}$.

On the submanifold ${\cal W}_e$ we will observe the deformation of the section ${\cal S}_e(u')$ as
a deviation of the geodesic congruence of $n^a$.
Especially the index theorem of the Morse theory\cite{MT} is powerful to see topological aspects.
Next, we mention the method to examine the limit of ${\cal S}_e$ near $\sc$. The following corollary \cite{Sorkin} of the Poincar\'e-Hopf's theorem is useful.

\newtheorem{theo}{Theorem}
\renewcommand{\thetheo}{}
\begin{theo}
Let M be a compact n-dimensional ($n>2$ is an odd number) $C^r(r\ge 1)$ manifold with $\Sigma_1\cup\Sigma_2=\dot{M}$ and $\Sigma_1\cap\Sigma_2=\emptyset$. X is any $C^{r-1}$ vector field with at most a finite number of zeros, satisfying the following two conditions: (a)The zeros of X are contained in Int M. (b)X has inward directions at $\Sigma_1$ and outward directions at $\Sigma_2$. Then the sum of the indices of X at all its zeros is related to the Euler numbers of $\Sigma_1$ and $\Sigma_2$:
\begin{equation}
\chi(\Sigma_2)-\chi(\Sigma_1)=2\ {\rm index}(X), \label{eq:theorem}
\end{equation}
\end{theo}    
where index$(X)$ is given by the alternating sum of the Morse number $\mu_k$ as $\rm{index}=\sum_k (-1)^k\mu_k$. The Morse number $\mu_k$ is the number of a critical point (zero of vector field $X$) whose index is $k$. The index of a critical point is given by the number of negative eigenvalue of Hesse matrix $H_{ab}=\nabla_a X_b$.  
This theorem means that when the topology of slices of a manifold changes, there must be a zero of the vector field on it and equation (\ref{eq:theorem}) is satisfied. 

Here we consider the case where an event horizon is spatial two spheres like the spacetime of C-metric.
As discussed above and illustrated in Fig. \ref{fig:3}, in this case $H_i$ approaches to $H^+$.
In other words, there is an embedding
\begin{equation}
\phi:{\cal W}_e\mapsto {\cal N},
\end{equation}
in which ${\cal N}$ is a trouser-like manifold with two boundaries $\Sigma_1=S^2$ and 
$\Sigma_2=S^2\cup S^2$.

\begin{figure}[htbp]
\begin{center}
\includegraphics[width=.60\linewidth]{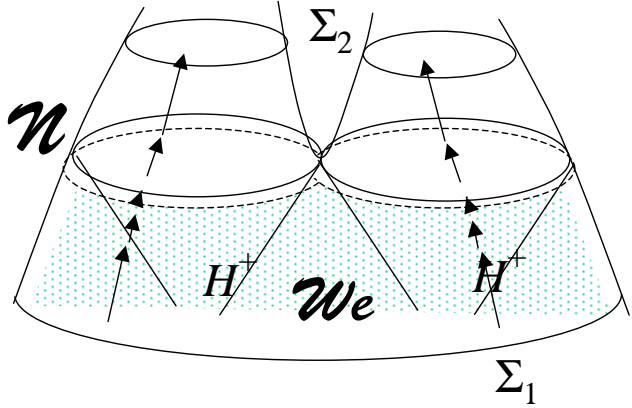}
\end{center}
\caption{Timelike submanifold ${\cal W}_e$ is diffeomorphically embedded into a trouser-like manifold ${\cal N}$.
On the boundary of ${\cal W}_e$ there are zeros of the tangent $n^{\prime a}$.}
\label{fig:4}
\end{figure}

 Because ${\cal W}_e$ and ${\cal N}$ are a three-dimensional 
manifold, we apply the theorem to $\cal N$. Let us consider a continuous tangent vector field $X$ of
$\cal N$ such that $X$ is the natural extension of $\phi^*(n^a)$
\begin{equation}
X^a|_{\phi({\cal W}_e)}=\phi_*(n^a),
\end{equation}
is outward directed on upper boundary $\Sigma_2$ and do not have a zero on ${\cal N}\setminus\overline{\phi({\cal W}_e)}$, since ${\cal N}\setminus\overline{\phi({\cal W}_e)}=\Sigma_2\times(-\infty,0]$.

 Here, we should pay attention to the fact that, in this case, zeros can exist only on the 
 boundary of ${\cal W}_e$, since ${\cal W}_e$ is complete about its own tangent $n^a$ by definition. 
As easily understood, the upper boundary of $\phi({\cal W}_e)$, $\partial(\phi({\cal W}_e))\setminus\Sigma_1$
is the set of endpoints of $X^a$ since $n^a$ is complete on ${\cal W}_e$.
Nevertheless, the endpoints (zeros) with index$=0$ do not affect the theorem 
and are extendible endpoints since Morse's lemma\cite{MT} states that the Morse critical point is generic.
By small reparametrization, we have $X^{\prime a}$ which is almost same as $X^a$ except for small 
neighborhood of $\partial(\phi({\cal W}_e))\setminus\Sigma_1$, and have only one zero on 
$\partial(\phi({\cal W}_e))\setminus\Sigma_1$ with index$=1$.

\begin{figure}[htbp]
\begin{center}
\includegraphics[width=.70\linewidth]{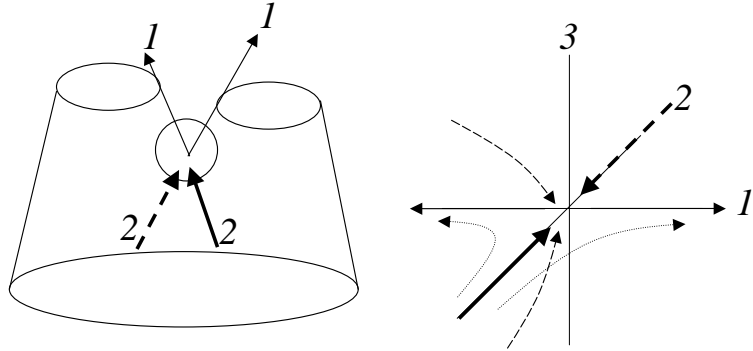}
\end{center}
\caption{The behavior of the vector field $n^a\equiv\tilde{g}^{ab}\tilde{\nabla}_b\Omega$ around the zero on a trouser-like
manifold. The left figure is $\cal N$ with one dimension suppressed and the zero is encircled.  
The right figure is close-up of the encircled point of the left figure. 
Each heavy arrow is in the direction of $x_2$-axis. 
Each dashed arrow lies on the $x_2x_3$-plane and each dotted arrow lies on the $x_1x_2$-plane. 
We should note that in the left figure $\cal N$ is drawn as two dimensional surface (actually it is a three dimensional surface) but that each right figure is as drawn the neighborhood of a zero on three dimensional surface. The region $\{(x_1,x_2,x_3|-x_1^2+x_2^2+x_3^2<0)\}$ is not contained in ${\cal W}_e$ and is the small extension in the direction of $X^a$. When we consider the expansion and the shear of the vector field $n^a$ along heavy arrows (the heavy arrow and the dotted heavy arrow), then we observe that approaching to the zero along heavy arrows, the shear becomes larger than the expansion.}
\label{fig:5}
\end{figure}

 Therefore we have to be careful to apply the theorem. Nevertheless, it would be possible to relate the index of the theorem and the congruence of null generators around the zero. From Morse's lemma, we suppose that the zero of the null generators is isolated. Since $\cal N$ is three 
 dimensions, the behavior of the congruence around the zero which is on ${\cal W}_e$ is illustrated in FIG.\ref{fig:5}.

The left figure in FIG.\ref{fig:5} is a three dimensional conformal embedding with one dimension suppressed. 
The right figure in FIG.\ref{fig:5} represents the close-up figure of encircled points of the left figure and makes clear the behavior in the vector field $X^a$ and $n^a$ in the neighborhood of their zero. 
We should  note that though in the left figure in FIG.\ref{fig:5} $\cal N$ is drawn as two dimensional surface rather than three dimensional surface, in the right figure the neighborhood of the zero is drawn as the neighborhood of the zero on three dimensional surface. 
In this figure, there is a zero with $\rm{index}=+1$. 
In the neighborhood of the zero, we can introduce a coordinate system $(x_1,x_2,x_3)$ on $\cal N$. 
A heavy arrow and a dashed heavy arrow are on $x_2$ axes, dashed arrows are on $x_2x_3$-planes, dotted arrows lie on $x_1x_2$-planes and solid arrows lie on $x_1x_3$-planes. 
In the left figure of FIG.\ref{fig:5}, the heavy arrow is on the front side  while the dashed heavy arrow is on its reverse side.

Seeing the region contained in $\phi({\cal W}_e)$, $\{(x_1,x_2,x_3)|-x_1^2+x_2^2+x_3^2>0\}$, we can extend its neighborhood into the other region in ${\cal N}\setminus\partial(\phi({\cal W}_e))$, $\{(x_1,x_2,x_3)|-x_1^2+x_2^2-x_3^2<0\}$. 
For it contains the region $\{(x_1,x_2,x_3)|-x_1^2+x_2^2+x_3^2>0\}$ but does not contain the other region $\{(x_1,x_2,x_3)|-x_1^2+x_2^2+x_3^2<0\}$ in a certain coordinate system. 
The limit behavior of the shear and the expansion of the vector field $n^a=\tilde{g}^{ab}\tilde{\nabla}_{b}\Omega$ tells us information about its zero since $\nabla_bn_a=\nabla_b\nabla_a\Omega$ gives Hesse matrix. 
Approaching to that zero along the heavy arrow, the dotted arrows on the $x_1x_2$-plane in FIG.\ref{fig:5} are going away from the zero, and dashed arrows on the $x_1x_3$-plane in FIG.\ref{fig:5} are approaching to the zero. 
Therefore, approaching to the zero along the heavy arrow in FIG.\ref{fig:5}, the shear of the vector field become larger than  the expansion.

Note that because at the zero with index$=1$ tangent vectors $X^a$ vanishes, it is Lifshitz continuous (that is, at least $C^{1-}$) at that point. Then, the embedding of ${\cal W}_e$ into $\cal N$ is $C^1$. 
We should note also that when we consider the embedding of a neighborhood of that zero, the into-mapping is not self-intersecting because of the fact we introduce a trivial patch with ${\bf R}^3$ in the neighborhood of that zero.  
Furthermore we comment that because in Schwarzshild spacetime, there are nothing but zeros with index$=0$ of vector field $X^a$ at the boundary of ${\cal W}_e$. ${\cal W}_e$ have a trivial completion into 
${\cal N}\sim S^2\times[0,1]$.

To summarize, for the Hesse matrix $H_{ab}=\tilde{\nabla}_an_b=\tilde{\nabla}_a\tilde{\nabla}_b\Omega$, we solve the eigenvalue equation
\begin{equation}
H_{ab}x^a=\Lambda x_b=\Lambda h_{ab}x^a
\end{equation}
where $h_{ab}$ is the metric of ${\cal W}_e$. Considering three basis vectors of ${\cal W}_e$, $\hat{n}^a=n^a/\sqrt{|n^cn_c|}, 
(m^a\pm\overline{m}^a)/\sqrt{2}$, the eigenvalues of $H_{ab}$ are given by

\begin{equation}
-\frac{n^an^b\nabla_an_b}{\sqrt{|n^cn_c|}},\ \ \mu\pm|\lambda|\ \ .
\end{equation}

In compact gauge, $n^{\prime c}n'_c$ is negative not on $\sc$ and converges to zero at $u'\rightarrow\infty$.
Therefore the first eigenvalue is negative and the index is determined by inequality $|\mu|<|\lambda|$.
In this gauge, $\Omega'=\omega\Omega$ is decreasing positive function about $u'$ and converges to 
zero in $u'\rightarrow\infty$.
Then we are able to choose two gauge conditions, they are $\partial_{u'}\omega/\omega\rightarrow 0$ and 
$\partial_{u'}\omega/\omega\rightarrow\beta<0$ and corresponds to $\omega\propto u^{\prime-\alpha},\ (\alpha>0)$ and
$\omega\propto e^{-\beta u'}$.
$\omega\propto e^{-\beta u'}$, however, means $n^an_a$ is not zero in $u'\rightarrow\infty$ and this 
is a bad aspent. When we take this gauge, the $n^a$-integral curves go into a timelike infinity.
Nevertheless the timelike infinitiy is generally pathological and $\overline{M}$ is not manifold there.
It will mislead topological investigations.

Therefore we must choose a gauge 
\begin{equation}
\omega=u^{\prime-\alpha}, \mu'=\frac{\alpha}{u'}, 
\label{eqn:gc}
\end{equation}
where higher order of $\mu'$ about $\Omega$ is eliminated by higher order gauge condition of $\omega$.

So if the behavior of Weyl curvature on $\sc$ is $\Psi_{4(1)}\sim u^{\prime-\epsilon}$ as $u'\ri\infty$,
$\lambda'$ behaves as $\lambda'\propto u^{\prime 1-\epsilon}\Omega'$ by eq.(\ref{eq:shear}).
The ratio of the shear to the expansion is 
\begin{equation}
\biggl|\f{\lambda'}{\mu'}\biggr|\sim u^{\prime 2-\epsilon}\O'+O(\O^{\prime 2}),
\end{equation}
From (\ref{eq:nn2})(\ref{eq:mupk})(\ref{eqn:gc}),
\begin{eqnarray}
0>n^{\prime c}n'_c&=&n^{\prime c}\tilde{\nabla}_c\Omega'=k_{(1)}\Omega'+...,\\
\Omega'&\propto& u^{\prime -2\alpha},\label{eq:ratio2}\\
\biggl|\f{\lambda'}{\mu'}\biggr|&\sim& u^{\prime 2-2\alpha-\epsilon},\label{eq:ratio}
\end{eqnarray}
where it should be emphasized that $\epsilon$ depends on gauge choicing $\omega$ and is a function of $\alpha$. 

%On the other hand, if we took a bad gauge condition $\omega=e^{-\beta u'}$, $\mu'$ is $\beta>0$.
%Then $\lambda'_{(1)}$ is given by $\lambda'_{(1)}=-e^{3\beta u'}\int\Psi_{4(1)}e^{-3\beta u''}du''$ and does not decay
% even
%for rapidly decaying $\Psi_{4(1)}\propto u^{\prime-\epsilon}$. 
%Nevertheless, since it is valid that $\Psi_{4(1)}$ decays in affine gauge $u$, $\Psi_{4(1)}\propto u^{-\epsilon}\sim 
%e^{-\epsilon\beta u'}$, $\lambda'_{(1)}$ decays and $\mu'$ dominates in this gauge.
%As stated above, this is a bad aspects of this gauge that a gauge make diverge the topological structure, since 
%the eternallypositive $\mu'$ implies diverging of the section ${\cal S}_e(u')$.

Therefore, if there is $\alpha$ satisfying $2-2\alpha-\epsilon>0$, this ratio can diverge, that is, the shear can be much larger than the expansion in this gauge. 
As shown above, though even the topological feature changed by the gauge condition it is guaranteed by the
existence of a gauge providing not decaying $\biggl|\f{\lambda'}{\mu'}\biggr|$ that the event horizon is topologically
non-trivial like FIG.\ref{fig:4}.

The point is that if  Weyl curvature falls off later than some power of parameter of tangent $n^a$ of $\sc$, two black holes will not coalesce eternally.
Now that we obtain the sufficient preparation, we can restate our sufficient condition for not coalescing black holes as follows;

\newtheorem{condition}{Condition}

\begin{condition}[Sufficient condition for separated two black holes]
When we transform the conformal factor from affine gauge into another gauge $\O\ri\O'=\O\o$, the number of black holes is not
only one, if there exists an unphysical spacetime $(\ti{M},\o^2\ti{g}_{ab})$ with the conformal factor $\o$ satisfying the below conditions.\\
(1)There is a real number $\alpha$ such that $\o=\o(\theta,\phi)u^{\prime -\alpha}$ and $\lim_{u\ri\infty}\o=0$, where $u$ and $u'$ are parameters of the null geodesics on $\sc$ in affine gauge and in another gauge, respectively and $\o(\theta.\phi)$ is the smooth function on a sphere which is positive and not singular everywhere.\\
(2)There is a real number $\alpha>0$ satisfying the following condition;
For any number $L>0$, there are positive numbers $K$ such that if $u'>K$, then $|u^{\prime 2-2\alpha-\epsilon}|>L$, where $\epsilon$ is an exponent appearing in $\Psi_{4(1)}$ extended in the power of $u'$, that is, $\Psi_{4(1)}\sim u^{\prime -\epsilon}$.
\end{condition}

Finally also we note that Einstein equation is only used through the definition of asymptotic flatness in the section 2 and Newman-Penrose equations. The condition 7 in Definition \ref{def:af} in the section 2 is equal to the condition that energy momentum tensor goes to zero at future null infinity rapidly if we impose the condition that Einstein equation holds. We used this condition in the eq.(\ref{eq:rev}) and so on. (Though in the book of Wald (reference \cite{Wald}) vacuum Einstein equation is assumed in the neighborhood of future null infinity, we weaken this requirement.) 
We never suppose something like energy conditions.

Since this condition is a sufficient condition, it is not clear whether the spacetime satisfying the condition is realistic.
In the next section, by proving in C-metric the (\ref{eq:ratio}) diverges ($\epsilon<2$) in appropriate gauge, we show that 
the event horizon is not that of a single black hole but that of two black hole connected at the virtual boundary of
conformally compactified manifold. This is what we already know about the causal structure of C-metric spacetime.

\section{\label{sec:5}C-metric as an concrete example}
To verify the relevance of the condition,
 we take an example which satisfies the above condition. Now we treat the vacuum C-metric \cite{Bonnor},\cite{Bonnor2} as an example of the spacetime in which two black holes never coalesce from the viewpoint of an observer outside 
black holes \cite{Cornish and Uttley}.
The line element of the vacuum C-metric is written in the form of
\beq
ds^2&=&-Hdu^2+H^{-1}dr^2+2Ar^2H^{-1}drdx\no\\
    &{}&+(G^{-1}+A^2r^2H^{-1})r^2dx^2+r^2Gd\varphi^2,\label{eq:cmetric}
\eeq
where 
\beq
H&=&1-2mr^{-1}+6Amx+ArG_{,x}-A^2r^2G,\\
G&=&1-x^2-2Amx^3.
\eeq
In BMS coordinate\cite{BMS}, it is expanded at large distance. 
Since in our discussion it is sufficient to consider the only neighborhood of $\sc$, we may transform the metric (\ref{eq:cmetric}) to a BMS coordinate written by

\beq
ds^2&=&-fdu^2-2e^{2\beta}dudr-2Udud\theta\no\\
    &{}&+r^2(e^{2\gamma}d\theta^2+e^{-2\gamma}\sin^2\theta d\phi^2),\label{eq:BMS}
\eeq
where four functions are 
\beq
f(u,r,\theta)      &=& 1-\f{2M(u,\theta)}{r}+O\bl(\f{1}{r^2}\br),\label{eq:f}\\
\beta(u,r,\theta)  &=& -\f{c^2(u,\theta)}{4r^2}+O\bl(\f{1}{r^3}\br),\label{eq:b}\\
\gamma(u,r,\theta) &=& \f{c(u,\theta)}{r}+O\bl(\f{1}{r^2}\br),\label{eq:g}\\
U(u,r,\theta)      &=& -(c_{\theta}+2c\cot\theta)+O\bl(\f{1}{r}\br),\label{eq:U}
\eeq
and $M(u,\theta)$ is given by
\begin{equation}
M_{,u}=-c_{,u}^2+\f{1}{2}(c_{,\theta\theta}+3c_{,\theta}\cot\theta -2c)_{,u}.
\end{equation}
We should note that these four functions are expressed by one function $c(u,\theta)$ whose differentiation by $u$ is called \textit{the news function},  
\begin{equation}
c_{,u}=\f{-3a^3m}{4A^3}\f{\sin^2\theta}{u^4}.
\label{eqn:nf}
\end{equation}

We introduce a new coordinate by
\beq
v\equiv u+2r.
\eeq
In this coordinate, the metric (\ref{eq:BMS}) becomes the following form,
\beq
ds^2&=&-(f-e^{2\beta})du^2-e^{2\beta}dudv-2Udud\theta\no\\
    &{}&+\f{1}{4}(v-u)^2(e^{2\gamma}d\theta^2+e^{-2\gamma}\sin^2\theta d\phi^2). 
\eeq
Furthermore, we introduce a new coordinate $V=1/v$ so that infinity along outgoing null geodesics will correspond to $V=0$. The metric components in the new coordinate $(u,V,\theta,\phi)$ are given by
\beq
ds^2&=&-(f-e^{2\beta})du^2+\f{e^{2\beta}}{V^2}dudV-2Udud\theta\no\\
    &{}&+\f{1}{4}(v-u)^2(e^{2\gamma}d\theta^2+e^{-2\gamma}\sin^2d\phi^2). 
\eeq 
In turn, we conformally transform the above physical metric into
\beq
\ti{ds^2}&=&-(f-e^{2\beta})\O^2du^2+e^{2\beta}dud\O-2U\O^2dud\theta\no\\
         &{}&+\f{1}{4}(1-\O u)^2(e^{2\gamma}d\theta^2+e^{-2\gamma}\sin^2\theta d\phi^2),\label{eq:conformalBMS}
\eeq
where this unphysical metric is related to the physical metric by $\ti{g}_{ab}=\O^2g_{ab}$ and $\O=V$.
The four functions (\ref{eq:f}),(\ref{eq:b}),(\ref{eq:g}) and (\ref{eq:U}) are expanded as
\beq
f     &=& 1-4M(u,\theta)\O+O(\O^2),\\
\beta &=& -c^2(u,\theta)\O^2+O(\O^3),\\
\gamma&=& 2c(u,\theta)\O+O(\O^2),\\
U     &=& -(c_{,\theta}+2c\cot\theta)+[-2c(c_{,\theta}+2c\cot\theta)\no\\
      &{}&+2(2c_{,u}+3cc_{,\theta}+4c^4\cot\theta)]\O+O(\O^2)
\eeq
in powers of conformal factor $\O$. 

Now let us determine the basis vectors defined in section \ref{sec:4} from the metric (\ref{eq:conformalBMS}). It is easy to decompose the metric (\ref{eq:conformalBMS}) into the form of
\begin{equation}
\ti{g}_{ab}=-(k_{(1)}\O+\cdots)l_al_b-2l_{(a}n_{b)}+2m_{(a}\bar{m}_{b)}.
\end{equation}
We see that the four dual vectors,
\beq
n_a&=&(d\O)_a,\label{eq:defn}\\
l_a&=&-\f{1}{2}e^{2\beta}(du)_a,\\
m_a&=&\f{1}{2\sq{2}}(1-u\O)\bl(-e^{-\gamma}\f{4U\O^2}{(1-u\O)^2}(du)_a\no\\
   &{}&+e^{\gamma}(d\theta)_a-ie^{-\gamma}\sin\theta(d\phi)_a \br),\\
\bar{m}_a&=& \f{1}{2\sqrt{2}}(1-u\O)\bl(-e^{-\gamma}\f{4U\O^2}{(1-u\O)^2}(du)_a\no\\
   &{}&+e^{\gamma}(d\theta)_a+ie^{-\gamma}\sin\theta(d\phi)_a \br),  
\eeq
satisfies the eauation (\ref{eq:omega}).

We raise the indices of the dual vectors to obtain four basis vectors,
\beq
n^a &=& 2e^{-2\beta}\bl(\f{\pa}{\pa u}\br)^a\no\\
    &+& [4(f-e^{2\beta})e^{-4\beta}\O^2+16e^{-4\beta-2\gamma}\f{U^2\O^4}{(1-u\O)^2}]\bl(\f{\pa}{\pa \O}\br)^a\no\\
    &+& 8e^{-2\gamma-2\beta}U\f{\O^2}{(1-u\O)^2}\bl(\f{\pa}{\pa \theta}\br)^a,\label{eq:n}\\
l^a&=&-\bl(\f{\pa}{\pa \O}\br)^a,\\
m^a &=&\f{\sq{2}e^{-\gamma}}{1-u\O}\bl(\f{\pa}{\pa \theta}\br)^a+i\f{\sq{2}e^{\gamma}}{\sin\theta(1-u\O)}\bl(\f{\pa}{\pa \phi}\br)^a,\\
\bar{m}^a &=&\f{\sq{2}e^{-\gamma}}{1-u\O}\bl(\f{\pa}{\pa \theta}\br)^a-i\f{\sq{2}e^{\gamma}}{\sin\theta(1-u\O)}\bl(\f{\pa}{\pa \phi}\br)^a.
\eeq
From (\ref{eq:defn}) and (\ref{eq:n}), we compute the norm of $n^a$ as follows.
\begin{equation}
n^an_a=4(f-e^{2\beta})e^{-4\beta}\O^2+16e^{-4\beta-2\gamma}\f{U^2\O^4}{(1-u\O)^2}.\label{eq:nn3}
\end{equation}
Since the first order of $\Omega$ in the equation (\ref{eq:nn3}) vanishes, we see that the choice of conformal factor $\O=V$ is affine gauge at $\sc$. That is, in the choice of this gauge
\begin{equation}
\mu_{(0)}=-p_{(0)}=-\f{1}{2}k_{(1)}=0
\end{equation}
holds. The shear $\lambda$ in this gauge is 
\beq
\lambda&\equiv& -\bar{m}^a\bar{m}^b\na_an_b\no\\
       &=&     -4c_{,u}\O+O(\O^2)\\
       &=&     \f{3a^3m}{A^3}\f{\sin^2\theta}{u^4}\O+O(\O^2),
\label{eqn:cmlm}
\eeq
where we used the fact the news function of the vacuum C-metric~\cite{Tomimatsu} is given by (\ref{eqn:nf})
and $u$ is the affine parameter of the null generator of $\sc$. \\
For the investigation in previous sections we require a gauge transformation from the affine gauge into the gauge such that $\sc$ can be compactified in future direction. Then we examine the ratio of the shear to the expansion of the congruence of $n^a$-integral curves which plunge into the zero at $n^a$-limit. In affine gauge, the shear in this direction behaves 
\begin{equation}
\lambda\sim\lambda_{(1)}\Omega\sim\f{1}{u^4}\O\label{eq:lamu},
\end{equation}
as $u\ri\infty$ and $\Omega\ri 0$.
 Under the transformation $\O\ri\O'=\o\O$, there is the relation between affine parameter and the parameter of the geodesic in new gauge as follows
\beq
u =  \f{u^{\prime \alpha+1}}{\alpha+1} & \mbox{($\alpha\ne -1$)}
\label{eq:uu}     
\eeq
\noindent
where we set $\o=u^{\prime-\alpha}\ \ (\alpha>0)$.\\
From (\ref{eq:mm})(\ref{eqn:gtlm0})(\ref{eq:lamlam})(\ref{eq:shear})(\ref{eq:ratio2})(\ref{eq:lamu}) and (\ref{eq:uu}), the ratio of the shear to the expansion is given by
\begin{equation}
\biggl|\f{\lambda'}{\mu'}\biggr|\sim u^{\prime\alpha+1}\O'\sim u^{\prime 1-\alpha}.
\end{equation}

Then, we see that when $u\ri\infty (u'\ri\infty)$, this ratio diverges to infinity for $0<\alpha<1$. We ensured the existence of the conformal factor $\O'=\o\O$ such that the $n^a$-limit ratio of the shear to the expansion of the $n^a$-congruence around $\sc$ diverges. We also find easily this gauge is compact gauge,

\begin{equation}
\lim_{u\ri\infty}\o=0.
\end{equation}
On the other hand, though Newman-Penrose formalism is ,of course, true, we confirm that the behavior of Weyl curvature is obtained directly from the metric
\beq
\tilde{\Psi}'_{4(1)}\sim\f{1}{u^{\prime 1-\alpha}}.
\eeq 
If we choose $\omega$ so that $0<\alpha<1$, $0<\epsilon<2$ holds. 
Therefore, two conditions (1) and (2) of the sufficient condition stated in the end of section 4 are satisfied. 
This implies that the two black holes of the C-metric are eternally separated.

\section{\label{sec:8}Summary and Discussions}

We have investigated the limit structure of the congruence of the $n^a$-integral curves and its orthogonal
spatial section introducing a pseudo-Newman Penrose basis vectors.
It was shown the sufficient condition in order that two black holes never coalesce to an asymptotic observer outside black holes in
such a spacetime with two black holes.
This is that if the tail of outgoing gravitational wave $\Psi_4$ remains at sufficiently late time for an asymptotic observer, 
two black holes remain separated forever.

In other words,
we clarify that the behavior of Weyl curvature $\Psi_{4}$ in the neighborhood of future null infinity at late time rules the geometrical structure of the congruence of tangent of $\sc$. 
Especially it is considered in spacetimes with two black holes, the two black holes separated forever is caused
by the shear of the congruence through the Poincar\'{e}-Hopf theorem. 
Moreover, we made sure that Weyl curvature of C-metric remained so late that it satisfy the sufficient condition.

Here we comment that to discuss the final stage of the gravitational collapse we have only consider causal structures
 in the
limit of the future direction instead of supposing stationary spacetime. Though at first sight they might correspond to each other, 
the authors know nothing about that. 
More advanced studies in the course of the present investigation will tell something about that.

Studying black hole perturbation in some exact solution background, it have been known that there is  gravitational
 wave tail in late time limit\cite{price}.
For example, it is easily confirmed that outgoing gravitational wave is far smaller than the late discussed in this article
for Schwarzschild spacetime. 
This is valid  since Schwarzschild spacetime is a single black hole spacetime.
We guess that also in other single black hole spacetimes the situation is same. 
Otherwise, the topology of black hole
will be changed by the effect of the gravitational wave perturbation.

As a future work we need investigate the physical and geometrical meaning of the parameter $u'$, althogh in this article, we treat it as the parameter such as future null infinity is compactified. It is interesting to investigate the relation between the compact gauge and the BMS group, however, at this stage, we do not know the answer.       

While we concentrate on the gravitational radiation $\Psi_{4}$ of Weyl curvature as leading order contribution, 
there are possibilities that other component of curvature may contribute to the strucutre of tangent $n^a$ of $\sc$
in higher order. For example, in Majumdar-Papapetrou spacetime, while gravitational field radiation $\Psi_{4}$ vanishes, we observe that $\Psi_{2}$
 contributes to the congruences of $n^a$ in a higher order.    

Furthermore, when we consider a higher dimensional black hole spacetime, a investigation like in the present article
will reveal the mechanism of topological black holes, since a higher dimensional counterpart of the C-metric spacetime
is black ring solution with vanishing angular momentum\cite{br}.

Using the result of this article, what can we know about the collapsing stars from the observation of the gravitational radiation ? 
The sufficient condition for separated black holes suggests the power of late time tail radiation will let us know 
about the final number of black holes. 
Of course, to get powerful statement about that, we should make clear the astrophysical meaning of the gauge choice 
used in this article. 

\section*{Acknowledgments}
We thank Dr. Nakamura for useful comments.
We would like to thank Professor A. Hosoya and T. Shiromizu for continuous encouragement. This work was supported by a 21st Century COE Program at
TokyoTech "Nanometer-Scale Quantum Physics" by the
Ministry of Education, Culture, Sports, Science and Technology.

\end{document}